\documentclass[ijoc]{informs4_supp}
\usepackage{eqndefns-left} 
\RequirePackage{tgtermes}
\RequirePackage{newtxtext}
\RequirePackage{newtxmath}
\RequirePackage{bm}
\RequirePackage{endnotes}

\OneAndAHalfSpacedXII 

\usepackage{algorithm}
\usepackage{algpseudocode}
\usepackage{tikz}
\usepackage[export]{adjustbox}
\usepackage{caption}
\usepackage[labelfont=sf]{subcaption}
\usepackage{todonotes}
\usepackage{soul}
\usepackage[normalem]{ulem}
\usepackage{amsmath,amssymb,amsfonts}
\usepackage{rotating}
\usepackage{fancyvrb}
\usepackage[table]{xcolor}
\usepackage{graphicx}
\usepackage[autostyle]{csquotes}
\renewcommand{\mkbegdispquote}[2]{\itshape}
\usepackage{hyperref}

\usepackage{fancyhdr}    

\fancypagestyle{titleheader}{
  \fancyhf{} 
  \fancyhead[C]{%
	\\[-8ex]%
	{\color{red}\small\textbf{Author-accepted manuscript. Published in INFORMS Journal on Computing: \url{https://doi.org/10.1287/ijoc.2024.1025}}}
  }
}

\long\def\fignote#1{\par\noindent
\FigureNoteStyle\HD{11}{0}{\it\FigureNoteName}\enskip #1\endgraf}

\newcommand{\sqboxs}{1.2ex}
\newcommand{\sqbox}[1]{\textcolor{#1}{\rule{\sqboxs}{\sqboxs}}}
\newcommand{\sqboxblack}[1]{\setlength{\fboxsep}{0pt}\fbox{\sqbox{#1}}}

\usepackage{natbib}
 \bibpunct[, ]{(}{)}{,}{a}{}{,}%
 \def\bibfont{\small}%
\EquationsNumberedThrough    

\TheoremsNumberedThrough     
\ECRepeatTheorems  %

\MANUSCRIPTNO{}

\begin{document}


\RUNAUTHOR{Gardeyn, Vanden Berghe and Wauters}

\RUNTITLE{An Open-Source Collision Detection Engine for 2D Irregular C\&P Problems}

\TITLE{Decoupling Geometry from Optimization in \break
     2D Irregular Cutting and Packing Problems:\break
     an Open-Source Collision Detection Engine}
\ARTICLEAUTHORS{%
     \AUTHOR{Jeroen Gardeyn}
     \AFF{KU Leuven, Department of Computer Science, Belgium, \EMAIL{jeroen.gardeyn@kuleuven.be}}

     \AUTHOR{Greet Vanden Berghe}
     \AFF{KU Leuven, Department of Computer Science, Belgium, \EMAIL{greet.vanden.berghe@kuleuven.be}}

     \AUTHOR{Tony Wauters}
     \AFF{KU Leuven, Department of Computer Science, Belgium, \EMAIL{tony.wauters@kuleuven.be}}
}

\ABSTRACT{%
     Addressing irregular cutting and packing (C\&P) optimization problems poses two distinct challenges: the geometric challenge of determining whether or not an item can be placed feasibly at a certain position, and the optimization challenge of finding a good solution according to some objective function.
     Until now, those tackling such problems have had to address both challenges simultaneously, requiring two distinct sets of expertise and a lot of research \& development effort.
     One way to lower this barrier is to decouple the two challenges. 
     In this paper we introduce a powerful collision detection engine (CDE) for 2D irregular C\&P problems which assumes full responsibility for the geometric challenge.  
     The CDE (i) allows users to focus with full confidence on their optimization challenge by abstracting geometry away and (ii) enables independent advances to propagate to all optimization algorithms built atop it.
     We present a set of core principles and design philosophies to model a general and adaptable CDE focused on maximizing performance, accuracy and robustness.
     These principles are accompanied by a concrete open-source implementation called \texttt{jagua-rs}.
     This paper together with its implementation serves as a catalyst for future advances in irregular C\&P problems by providing a solid foundation which can either be used as it currently exists or be further improved upon.
}

\FUNDING{This research was supported by the Research Foundation --- Flanders (FWO) under grant number 1S71222N and K804824N.}




\maketitle
\thispagestyle{titleheader}  

\section{Introduction} \label{section:introduction}
2D cutting and packing (C\&P) problems involve placing a set of smaller items within the bounds of a larger container.
This broad class of optimization problems can be very challenging to solve, especially when the items or containers are \textit{irregular} (non-rectangular) in shape.
This difficulty has led to a highly active yet fractured research domain featuring many approaches to many problem variants.

While the objectives and particulars of these irregular C\&P --- \textit{nesting} --- problems can differ, they all share the need for a specific feasibility check: determining whether or not an item can be cut from or placed at a certain position without overlapping with anything else.
Unlike, say, the traveling salesman problem in which feasibility simply means ensuring all locations are visited, here the irregularity of the items and containers makes implementing this check in a precise and efficient manner a challenging endeavor.
Therefore, in addition to the \textbf{optimization challenge} of finding a good solution according to some objective function, irregular C\&P problems also pose an additional \textbf{geometric challenge} of accurately and efficiently verifying the feasibility of candidate solutions.

\begin{figure}[h]
     \FIGURE
     {\includegraphics[width=1\linewidth]{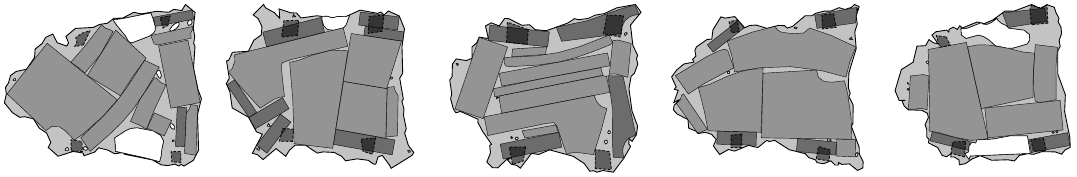}}
     {Example solution of a leather nesting instance. \label{fig:example_irr_cp}}
     {Dataset introduced by \citet{baldacci2014algorithms}.}
\end{figure}

By way of example, let us consider the leather nesting problem depicted in Figure \ref{fig:example_irr_cp}.
The aim is to use the leather hides as efficiently as possible (the optimization challenge), while ensuring that all items are fully contained within the hides and are not colliding with each other or any defects (the geometric challenge).
When a human is tasked to solve such a problem, they can rely on their innate spatial reasoning skills to ensure no \textit{collisions} occur and even subconsciously rule out many invalid placements before ever considering them.
However, translating this particular human intuition into a robust and efficient algorithm is far from trivial.
The general problem of collision detection also presents a significant challenge in other domains, such as computer graphics and simulations.
Indeed, this seemingly simple problem is so challenging that entire books have been dedicated to the subject (\citet{ericson2005real} for example). 

From a conceptual and algorithmic perspective, the geometric and optimization aspects of such problems are clearly distinct:
the feasibility of a solution implies nothing about its quality or the strategy used to generate it.
At present, however, there are no publicly available tools to handle collision detection in irregular 2D C\&P problems, meaning that researchers and practitioners need to simultaneously address the geometric and the optimization challenges.
This is an inherently inefficient situation that forces researchers to spend a significant proportion of their time reinventing the wheel.
Not only do they need to address the optimization intricacies of their new 2D irregular C\&P variant or the methodological challenges of developing a new approach for an existing problem, but they also need to spend effort integrating this with a feasibility check despite the fact that this geometric challenge is a feature of all problem variants.
Besides the wasted effort, this also creates a significant barrier to entry that impedes further advances in the field.

In this paper, we will decouple the geometric from the optimization challenge by introducing an open-source, easy-to-use and adaptable \textit{collision detection engine} (CDE) that is capable of handling the geometric component of 2D irregular C\&P problems.
The primary aim of the CDE is to (in)validate potential item placements as efficiently and accurately as possible.

We envisage two target audiences for this tool.
First, there are those who simply want to focus on the combinatorial aspects of their 2D irregular C\&P problem at hand.
They stand to benefit greatly from having a freely available engine to incorporate into their own methodology, which essentially outsources the geometric challenge and enables them to focus their efforts on developing smart optimization algorithms.
The second target audience are those who, rather than solving optimization problems themselves, might have good ideas concerning how to further refine or extend this open-source project. Our CDE will therefore not represent an endpoint, but should instead serve as the starting point for these two target audiences: either a tool to be readily used or one to be further improved upon.

The remainder of this paper is structured as follows.
In Section \ref{section:general_approaches} we review the general approaches to collision detection, highlighting their strengths and limitations.
The main body of the paper is then dedicated to the components of the engine itself, outlined in Section \ref{section:collision_detection_engine}, followed by a set of core principles and design philosophies which are each time accompanied by concrete implementations in Sections \ref{section:polygon_collision}-\ref{section:polygon_simplification}.
Finally, Sections \ref{section:codebase}-\ref{section:experiments} introduce our open-source CDE and discuss the factors that affect its performance.

\section{General Approaches to Geometry in 2D C\&P Problems} \label{section:general_approaches}
The literature contains many approaches to deal with the geometric challenge of 2D irregular C\&P problems.
The terms \textit{collide}, \textit{overlap} and \textit{intersect} are used interchangeably in the academic literature, but all refer to the same concept: two entities occupying the same space.
This paper will primarily use the word \textit{collision}.

A valuable introduction to geometry in irregular 2D C\&P and an overview of the most frequently used methods is provided by \citet{bennell2008geometry}.
In the interest of brevity, we will not exhaustively repeat the content of the tutorial.
Instead, we provide a short summary of the most relevant approaches reviewed by \citet{bennell2008geometry} along with a discussion of their main strengths and weaknesses.
The three most common approaches for dealing with geometry in irregular C\&P problems are: \textbf{raster}, \textbf{no-fit polygon} and \textbf{trigonometry}.

The simplest method is arguably the raster approach, where irregular shapes are discretized and represented as a set of pixels within a grid.
Collision detection is performed by simply checking the occupation of the pixels.
This strategy is flexible, robust and relatively straightforward to implement efficiently.
To guarantee feasibility, discretizing an irregular shape into a set of pixels inevitably requires containers to be deflated and items to be inflated.
As a result, a low grid resolution can severely affect the maximum achievable solution quality.
However, increasing the resolution is accompanied by a significant rise in computational cost and memory usage.
Hence, determining a suitable grid resolution is a delicate balancing act.
Furthermore, each possible rotation of an item requires a distinct discretization, which can become problematic when continuous rotation of items is allowed.

A second commonly used approach is the no-fit polygon (NFP), which encompasses the set of locations where two polygons will collide. A unique NFP must be generated for each pair of polygons and every possible rotation.
Detecting collisions between two polygons entails checking whether or not a single reference point of one of the polygons lies inside the respective NFP.
Section \ref{section:polygon_collision} will clarify why this is a relatively fast operation.
Additionally, the boundary of the NFP defines where two polygons touch each other.
This is very valuable information in an optimization context, where compact solutions are likely being targeted.
Similar to the raster approach, NFPs are not particularly suitable when rotation freedom is high.
For $n$ polygons with $m$ possible rotations, $(m*n)^2$ unique NFPs need to be computed.

The most significant challenge for this approach undoubtably remains the development of robust NFP generators, particularly for non-convex and complex shapes.
Over the years, numerous papers have been published claiming to improve NFP generation robustness or efficiency.
Yet, at the time of writing this article, a publicly available implementation without significant shortcomings remains elusive.
Even the authors of the most popular NFP generator on GitHub (\citet{github_nfp}) acknowledge that their implementation is not feature-complete and only works well for shapes without holes or concavities.

The third general approach for dealing with geometry in irregular C\&P problems is trigonometry.
Its precision, flexibility and simplicity make it an attractive choice.
However, approaches that rely directly on trigonometry are often quickly written off as being too slow for practical use.
For example, \citet{bennell2008geometry}, state:
\begin{displayquote}
     ``Hence, [trigonometry] does not lend
     itself to nesting algorithms based on any iterative
     search process since the time spent on evaluating
     the geometry can restrict the time, and therefore
     the extent, of the search. However, constructive
     algorithms, for instance approaches that build solutions based on a previously defined sequence of
     pieces, may quite efficiently use [trigonometry] to tackle
     the geometric issues of nesting problems.''
\end{displayquote}
The general claim being made here is certainly not incorrect.
By default, trigonometry is slow and highly sensitive to the number of edges in a polygon and therefore seldom employed.
For this reason, most state-of-the-art solution strategies for irregular C\&P problems are either based on a raster or on NFPs.

Another approach discussed by \cite{bennell2008geometry}, the phi-function, also deserves to be mentioned. 
While interesting, the limited accessibility of this method and lack of general phi-function `generators' have hindered its widespread adoption.

A common theme across all approaches is the severe lack of publicly available source code of the implementations.
This impedes scientific progress and leads to a lot of wasted time and effort.
We are determined to change the status quo and are guided by the principle of open science.

\begin{table}[h]
     \TABLE
     {Strengths and weaknesses of the general approaches for collision detection in 2D irregular C\&P problems. \label{table:state-of-the-art}}
     {\begin{tabular}{c|c c c}
               \hline
               Approach     & Precision         & Robustness        & Implementation ease$^*$ \\ \hline
               Raster       & $\star$           & $\star\star\star$ & $\star\star\star$       \\
               NFP          & $\star\star\star$ & $\star$           & $\star\star$            \\
               Trigonometry & $\star\star\star$ & $\star\star\star$ & $\star$                 \\ \hline
          \end{tabular}}
     {$^*$to reach a sufficiently fast implementation.}
\end{table}

Table \ref{table:state-of-the-art} presents the strengths and weaknesses of the three approaches discussed based on the criteria of precision, robustness and effort required to reach a sufficiently fast implementation.
Deciding on which approach to use as the basis of our implementation is not straightforward as no approach is without shortcomings.
However, given that the primary aim of our CDE is for its users to trust that they can safely delegate the geometric challenge, maximum precision and robustness are crucial.
Trigonometry is the only approach which does not compromise with respect these two properties.
Furthermore, increased implementation effort is a manageable drawback.
The open-source nature of our proposed implementation means that the bulk of the development the effort can be shared and reused.
Additionally, improvements in computational efficiency can easily be quantified, allowing for incremental improvements.

\section{Collision Detection Engine} \label{section:collision_detection_engine}
When solving irregular C\&P problems, the computational bottleneck will typically pertain to validating potential item placements, regardless of the search strategy one uses to improve the solution quality.
The speed at which these feasibility checks can be performed therefore inherently constrains the design of any optimization algorithm.
However, whether a solution is feasible or not implies nothing about its quality or the optimization strategy used to generate it.
Decoupling these two challenges by way of a reusable and extendable \textit{engine} to handle the geometry would enable researchers to focus with full confidence on the optimization challenge.

\begin{figure}[h]
     \FIGURE
     {\includegraphics[width=0.3\linewidth]{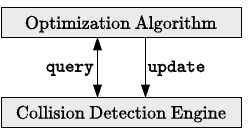}}
     {Fundamental interaction between the CDE and an optimization algorithm for irregular C\&P problems. \label{fig:diagram_cde_heur}}
     {}
\end{figure}

Figure \ref{fig:diagram_cde_heur} illustrates how a collision detection engine (CDE) fits into the process of solving a 2D irregular C\&P problem.
The CDE enables the geometric aspect of the problem to be abstracted away by assuming the responsibility of validating item placements.
To perform this task, two fundamental interactions between the CDE and the optimization algorithm must take place.

First, the CDE must be able to resolve \textbf{queries} by definitively answering whether or not an item can be placed at a certain position without causing a collision.
The interior of every item is represented by a base shape ($\mathcal{S}$), with every placement of this item corresponding to a transformation of its base shape ($\mathcal{S}^T$).
A placement is defined by a \textit{rigid transformation}, which is a combination of (i) a translation along the $x$ and $y$ dimensions, (ii) a rotation by angle $\theta$ and (iii) a reflection (or flip).
Determining if a placement is feasible involves checking if $\mathcal{S}^T$ is situated entirely inside the container and does not collide with any other items.

The second interaction concerns changes to the environment (for example placing or removing items).
These changes must be communicated to the CDE so it can \textbf{update} its internal state to reflect the new environment.
In practice, more sophisticated or tightly integrated interactions are possible, but Figure \ref{fig:diagram_cde_heur} outlines the bare minimum of interaction required.

Although collision detection also plays a central role in various other domains (video games, computer graphics, simulation), we cannot simply rely on any existing off-the-shelf engines.
To achieve maximum performance, these engines are typically tailored to the specific needs and unique characteristics of their respective domain.
By focusing on irregular C\&P problems, we can make a number of assumptions which will enable us to design a specialized CDE.

First, we can assume a dense environment (unlike video games), which means that the probability of collision is high.
Second, we can assume a largely static environment (unlike simulations), which means that queries will be more frequent than updates.
Third, we are only interested in the boolean answer to the query.
Therefore, we only need to determine whether a collision occurs, but not necessarily the details concerning where and how exactly it occurs.

\section{Polygon Collision} \label{section:polygon_collision}
At a fundamental level, detecting collisions entails determining whether or not two arbitrary polygons collide.
More specifically, we aim to detect collisions between \textbf{simple polygons}: polygons that do not intersect themselves and have no holes.
Convexity of the polygons is not a requirement.
Given a simple polygon $\alpha$, let us define some notation:
\begin{equation}
\begin{aligned}
     \mathcal{S}_\alpha       & : \text{set of all points in the interior of } \alpha \\
     E_\alpha = \{...,e,...\} & : \text{set of edges of $\alpha$}                     \\
     e                        & : \text{set of all points on the corresponding edge}
\end{aligned}
\end{equation}
Two shapes $\mathcal{S}_\alpha$ and $\mathcal{S}_\beta$ collide if a point $p$ exists which is in both $\mathcal{S}_\alpha$ and $\mathcal{S}_\beta$.
If no such point exists, the polygons do not collide:
\begin{equation}
     \mathcal{S}_\alpha \cap \mathcal{S}_\beta = \emptyset
     \iff
     \nexists p : p \in \mathcal{S}_\alpha \;\boldsymbol{\land}\; p \in \mathcal{S}_\beta
     \label{eq:poly_col_1}
\end{equation}
The remainder of this section describes a trigonometric approach for detecting collisions between two polygons.

\begin{figure}[ht]
     \FIGURE
     {
          \subcaptionbox{\label{fig:poly_coll_edges}}
          {\includegraphics*[width=0.25\textwidth]{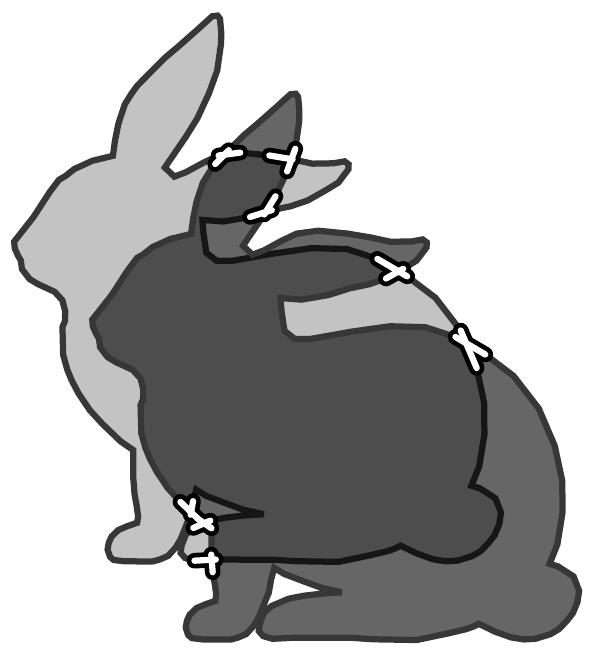}}
          \subcaptionbox{\label{fig:poly_coll_inclusion}}
          {\includegraphics*[width=0.25\textwidth]{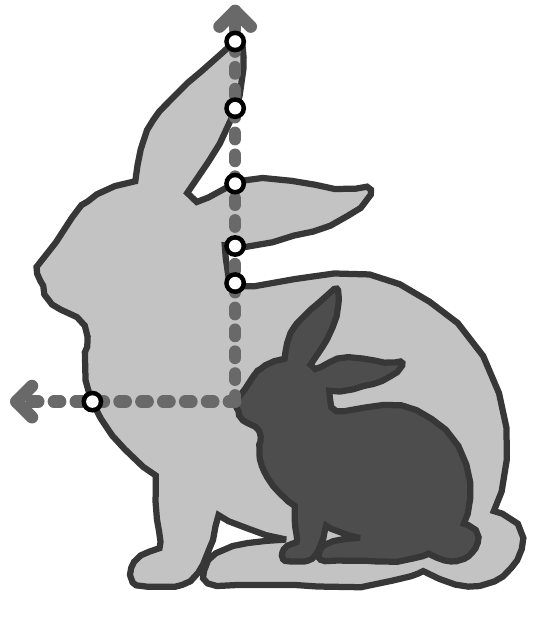}}
     }
     {Types of collision between a pair of simple polygons. \label{fig:polygon_collision}}
     {Collision due to (a) intersecting edges or (b) inclusion.}
\end{figure}

Figure \ref{fig:polygon_collision} illustrates the distinct types of collision that exist between two simple polygons.
The first type of collision occurs when the polygons have edges that intersect each other.
If there are no intersecting edges, the two polygons are either disjoint or one polygon is located entirely inside the bounds of the other.
The second type of collision accounts for the second case, where one polygon is fully contained within the other.
Every collision between two simple polygons can be expressed as one of these two types \cite{preparata1985computational}.

Figure \ref{fig:poly_coll_edges} shows an example of collision due to intersecting edges, where the intersecting edges are highlighted.
To detect this type of collision, all unique pairs of edges between both polygons need to be checked with a \textbf{line segment intersection} test, an efficient implementation of which was provided by \citet{antonio1992faster}.
Let us now define $\mathcal{E}_\alpha$ as the set of all points that lie on the boundary of simple polygon $\alpha$:
\begin{equation}
     \mathcal{E}_\alpha : \bigcup_{e \in E_\alpha} e
\end{equation}
Collision due to intersecting edges does not occur if none of the edges defining one polygon intersect with any of the edges defining another:
\begin{equation}
     \mathcal{E}_\alpha \cap \mathcal{E}_\beta = \emptyset
     \iff
     \forall e_\alpha \in E_\alpha, \forall e_\beta \in E_\beta : e_\alpha \cap e_\beta = \emptyset
\end{equation}
Proving there is no collision by intersecting edges involves checking all unique pairs of edges.
The computational cost of this proof scales quadratically with the number of edges: $O(|E_\alpha||E_\beta|)$.
In the example depicted in Figure \ref{fig:poly_coll_edges} each shape has 140 edges, resulting in a total of 19600 intersection checks required.
Despite these shapes obviously colliding, it is worth noting that only a very small minority of the edge-pairs ($\sim0.08\%$) actually intersect.

An example of collision due to inclusion is illustrated in Figure \ref{fig:poly_coll_inclusion}.
It shows a smaller shape $\mathcal{S}_\mu$ colliding with a larger one $\mathcal{S}_\Omega$ without any of their edges intersecting.
This is only possible when the smaller shape is fully enclosed within the larger shape ($\mathcal{S}_\mu \subseteq \mathcal{S}_\Omega$).
Determining that no inclusion occurs thus requires demonstrating that there exists a point $p \in \mathcal{S}_\mu$ which is $\not\in \mathcal{S}_\Omega$:
\begin{equation}
     \mathcal{S}_\mu \not\subseteq \mathcal{S}_\Omega
     \iff
     \exists p \in \mathcal{S}_\mu: p \notin \mathcal{S}_\Omega
\end{equation}
This can be achieved by performing a \textbf{point in polygon} test, with one approach involving the ray-casting algorithm introduced by \citet{shimrat1962algorithm}.
In Figure \ref{fig:poly_coll_inclusion}, two example rays are cast from a point in $\mathcal{S}_\mu$ to outside $\mathcal{S}_\Omega$.
The number of intersections between the ray and the edges of $\mathcal{S}_\Omega$ indicate whether or not $p \in \mathcal{S}_\Omega$.
An odd number of intersections means $p \in \mathcal{S}_\Omega$, while an even number of intersections proves $p \notin \mathcal{S}_\Omega$.
In the example both rays intersect $\mathcal{S}_\Omega$ an odd number of times (1 and 5 times).

Since only cases of full inclusion remain undetected by the edge intersection test, checking a single point is sufficient to prove no inclusion occurs.
Therefore, the computational cost of the inclusion test scales linearly with the number of edges of $\mathcal{S}_\Omega$.
In this example, a single point in polygon test requires 140 line segment intersection tests.
In practice, however, either shape could assume the role of $\mathcal{S}_\mu$ or $\mathcal{S}_\Omega$.
Given that this test is noncommutative, proving there is no inclusion consequently involves two point-in-polygon tests.

In summary, two polygons do not collide if none of their edges intersect and neither polygon is fully positioned inside the other:
\begin{equation}
     \mathcal{S}_\alpha \cap \mathcal{S}_\beta = \emptyset
     \iff
     \mathcal{E}_\alpha \cap \mathcal{E}_\beta = \emptyset
     \boldsymbol{\;\land\;}
     \mathcal{S}_\alpha \not\subseteq \mathcal{S}_\beta
     \boldsymbol{\;\land\;}
     \mathcal{S}_\beta \not\subseteq \mathcal{S}_\alpha
\end{equation}
This is a fundamentally precise and robust way of detecting collisions between any two polygons, but computational efficiency is lacking.
For the example depicted in Figure \ref{fig:polygon_collision}, where the shapes each have 140 edges, a naive implementation would require a total of 19880 line segment intersection tests (19600 + 140 + 140) to prove no collision occurs.
For the edge intersection test, more efficient algorithms exist. One example is by \citet{bentley1979algorithms}, which lowers the computational cost from a quadratic to quasi-linear time complexity.
When considering potentially hundreds of polygons, however, checking individual polygons against one another still represents an unacceptable computational burden.
The following sections will therefore investigate how we can improve the efficiency of this approach while retaining its inherent precision and robustness.

\section{Hazards} \label{section:hazards}
For the CDE to be easily adaptable, it should be agnostic of the specific problem variant being addressed.
We therefore need a unifying concept with which to express anything that could cause a collision and render the placement of an item invalid.
In this section, we introduce the concept of \textit{hazards}: an abstraction which is capable of expressing spatial constraints in a general way.

\begin{figure}[h]
     \definecolor{darkgray}{RGB}{22,22,22}
     \definecolor{lightgray}{RGB}{112,112,112}
     \FIGURE
     {\includegraphics[width=0.3\linewidth]{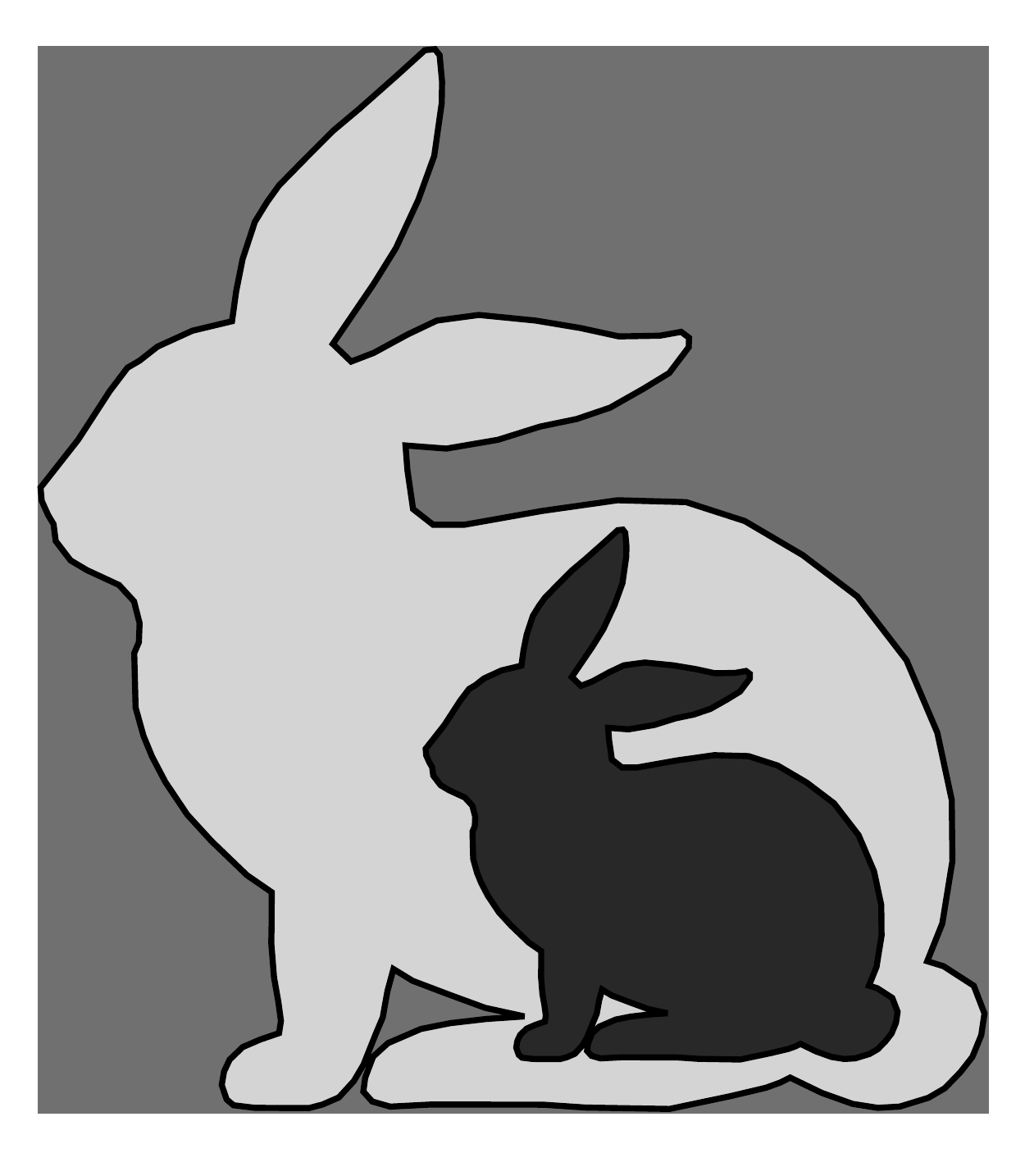}}
     {Different types of hazards. \label{fig:hazards}}
     {Two distinct hazards are present. 
     The container induces an \sqboxblack{lightgray} exterior hazard and the item induces an \sqboxblack{darkgray} interior hazard.}
\end{figure}

Figure \ref{fig:hazards} demonstrates the concept of hazards with an example of a container (the large bunny) inside of which an item (the smaller bunny) is placed.
Two distinct hazardous regions are present: one induced by the container and another induced by the item.
Since items must be fully situated inside the container, the hazard induced by the container refers to all of its \textbf{exterior} locations.
Meanwhile, the hazard induced by the item refers to every location within its \textbf{interior}.

In summary, the shape of hazard $h$ induced by entity $e$ is defined as:
\begin{equation}
     \mathcal{S}_h =
     \begin{array}{l}
          \left\{
          \begin{array}{lll}
               \mathcal{S}_e               & \text{if $e$ induces an interior hazard}
               \\
               {\mathcal{S}_e}^\complement & \text{if $e$ induces an exterior hazard}
          \end{array}
          \right.
     \end{array}
     \label{eq:hazard_shape}
\end{equation}
In case of interior hazards, the shape of the hazard is identical to the entity that induces it (an item, for example).
For exterior hazards, its shape is the complement of the entity which induces it. Placing an item $i$ with transformed shape $\mathcal{S}_i^T$ into a container with hazard set $H$ is feasible if it does not collide with any of the existing hazards:
\begin{equation}
     \mathcal{S}_H
     =
     \bigcup_{h \in H} \mathcal{S}_h
\end{equation}
\begin{equation}
     \mathcal{S}_i^T \cap \mathcal{S}_H = \emptyset
     \iff
     \forall h \in H : \mathcal{S}_i^T \cap \mathcal{S}_h = \emptyset
\end{equation}
The concept of hazards can be further expanded to cover additional constraints regarding the validity of item placements.
A quality zone for example, as first explored by \citet{heistermann1995nesting}, is a part of the container where the quality differs (lower or higher) from the rest.
Such zones can easily be represented as hazards, where zones with inferior or superior quality are converted into interior and exterior hazards, respectively.
In a problem context where each item $i$ must be produced from a minimum quality level, not all hazards are universally applicable.
When validating a placement, all hazards induced by higher-quality zones than the item requires may safely be ignored and only those induced by lower quality regions need to be taken into account. 
In this context, each item $i$ thus has its own set of \textit{relevant} hazards: $H_i \subseteq H$.

\section{CDE Implementation} \label{section:implementation_cde}

\begin{figure}[ht]
     \FIGURE
     {\includegraphics[width=0.5\linewidth]{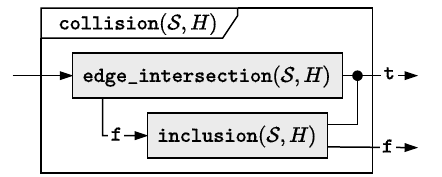}}
     {The \texttt{collision} function. \label{fig:diagram_collision}}
     {Returns whether shape $\mathcal{S}$ collides with any hazards $h \in H$, \texttt{t}: true, \texttt{f}: false.}
\end{figure}

The state of the CDE is defined by its set of registered hazards ($H$).
Figure \ref{fig:diagram_collision} provides a diagram of the \texttt{collision} function, which is used to query the CDE.
When validating the placement of an item $i$, its transformed shape ($\mathcal{S}_i^T$) can be passed to this function to determine if it collides with any hazard $h \in H$.

The remainder of this section details how we implemented the CDE and highlights the interactions between its various components.
All efficiency enhancements introduced throughout this section yield results that are identical to those produced by the naive implementation from Section \ref{section:polygon_collision}.

\subsection{Two-phased approach}

Many collision detection techniques in computer graphics and simulations employ a two-phased approach, which splits a procedure into a broad and a narrow phase.
The broad phase is responsible for eliminating the bulk of the work, using relatively fast algorithms.
The remaining work, which could not be inexpensively eliminated, is then resolved using time-consuming algorithms within a narrow phase.

\begin{figure}[h]
     \FIGURE
     {\includegraphics[width=0.4\linewidth]{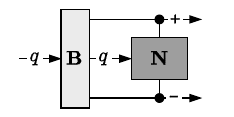}}
     {An abstract representation of the two-phased approach.\label{fig:diagram_two_phases}}
     {$\mathbf{B}$: broad phase, $\mathbf{N}$: narrow phase, $q$: query.}
\end{figure}

Figure \ref{fig:diagram_two_phases} illustrates the interaction between these two phases.
Assume we have an abstract query ($q$) which can be resolved to either a positive ($\boldsymbol{+}$) or negative ($\boldsymbol{-}$) answer.
The query is first sent through the broad phase ($\mathbf{B}$), which is often able to swiftly resolve it.
However, the quick checks employed by the broad phase are not always able to provide a definitive answer.
In such cases the query must be passed on to the narrow phase ($\mathbf{N}$), which is guaranteed to produce a definitive answer, albeit at a higher computational cost.

Section \ref{section:polygon_collision} already outlined a set of checks which can assume the role of the narrow phase.
What remains outstanding is designing a broad phase that focuses on maximizing overall performance.
This broad phase should strike the right balance between the ratio of queries it can resolve and its associated average computation cost.
The following sections will explore the two-phased approach in the context of the \texttt{edge\_intersection} and \texttt{inclusion} procedures from Figure \ref{fig:diagram_collision}.

\subsection{Quadtree}
The broad phase of our CDE revolves around a data structure based on quadtrees to represent the hazards.
Introduced by \citet{finkel1974quad}, quadtrees are tree-based data structures which recursively divide a two-dimensional space into quadrants.
They are used in computer graphics to efficiently store and query spatial data, with a practical example being image compression (\citet{shusterman1994image}).
Their efficiency stems from the fact that they provide high resolution only in areas where it is needed and do not waste any resources in regions where it is not required.

\begin{figure}[ht]
     \definecolor{white}{RGB}{255,255,255}
      \definecolor{gray}{RGB}{178,178,178}
      \definecolor{dark}{RGB}{54,54,54}

      \def\subfigwidth{.22\textwidth}
      \def\figwidth{\linewidth}
     \FIGURE
     {
          \subcaptionbox{$d=1$}
          {\includegraphics[width=\subfigwidth]{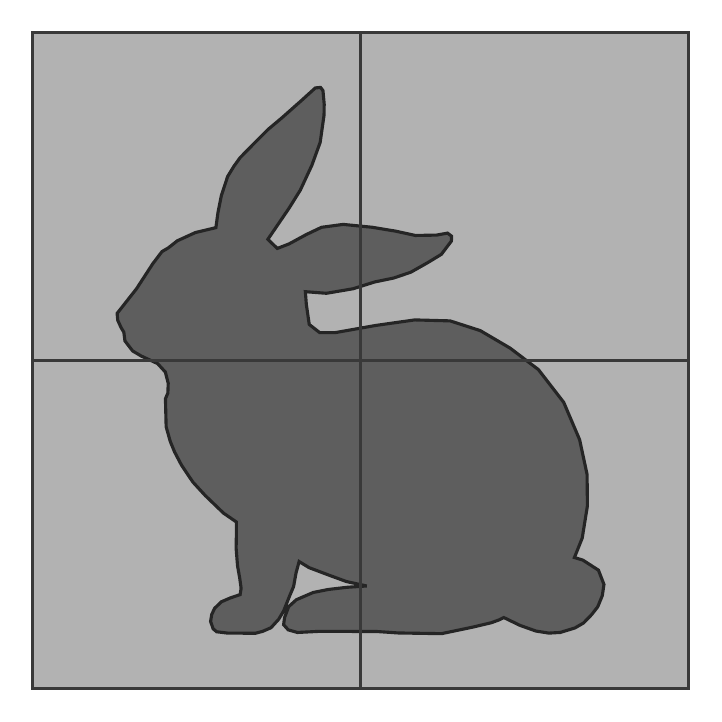}}
          \subcaptionbox{$d=2$}
          {\includegraphics[width=\subfigwidth]{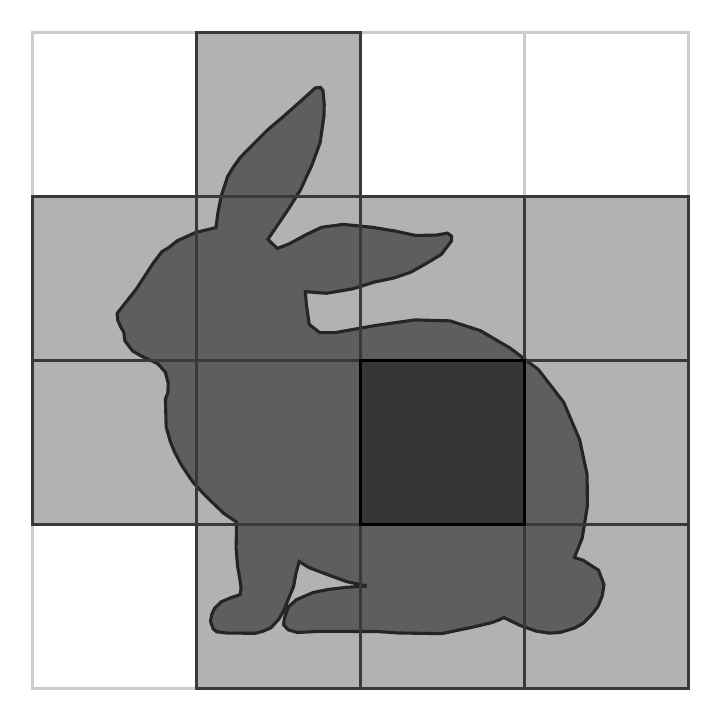}}
          \subcaptionbox{$d=3$}
          {\includegraphics[width=\subfigwidth]{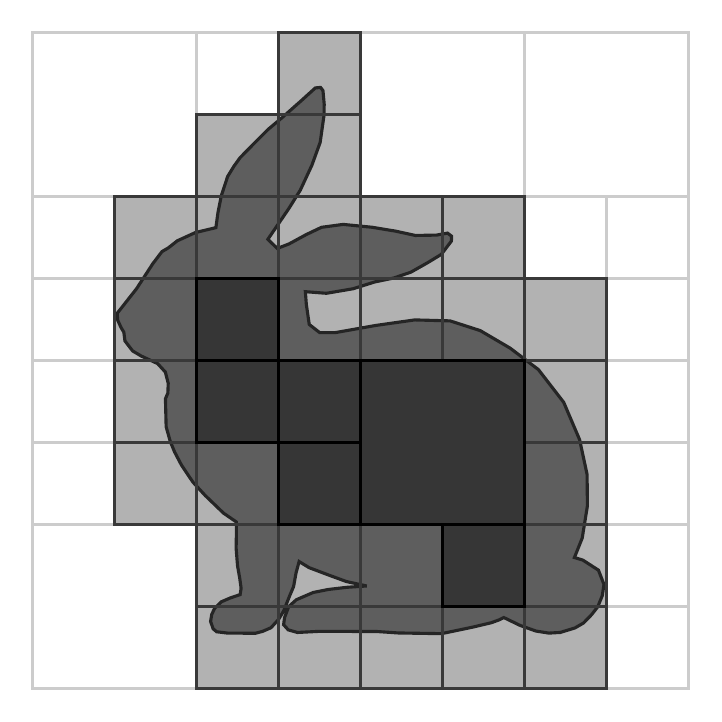}}
          \subcaptionbox{$d=4$}
          {\includegraphics[width=\subfigwidth]{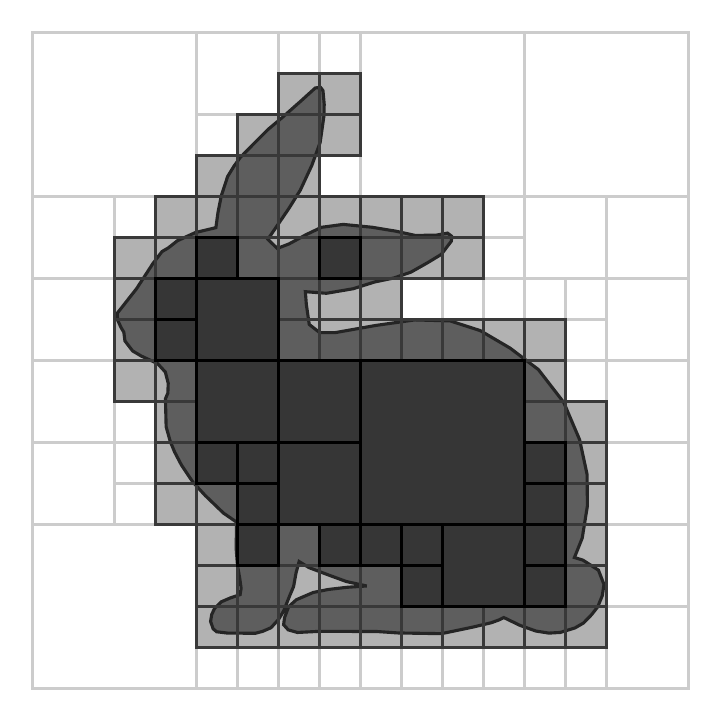}}
     }
     {Quadtree for an interior hazard with varying maximum depths: $d$. \label{fig:quadtree_generation}}
     {Hazard is $[$ \sqboxblack{white} not, \sqboxblack{gray} partially, \sqboxblack{dark} fully $]$ present in the node.}
\end{figure}

Figure \ref{fig:quadtree_generation} illustrates a quadtree for a single (interior) hazard.
The four illustrations depict the data structure with varying maximum tree depths ($d$).
Only the leaf nodes of the tree are visualized, and their colors correspond to the degree to which a hazard is present.
As depth increases, only nodes which partially contain hazards are subdivided into four children (if $< d$).
Nodes that do not feature any hazards or those fully occupied by them do not require any additional resolution provided by the extra depth.
In case of partially present hazards, additional information such as which edges of the hazard intersect with the respective node are also stored.
Although not shown in Figure \ref{fig:quadtree_generation}, it is important to note that a single node can contain information concerning multiple hazards.

The quadtree can be queried to efficiently narrow down which nodes (and hazards) are relevant for a given query.
The quadtree is part of the internal state of the CDE and must therefore be updated whenever the hazards change (such as when inserting or removing items). Both of these procedures will be discussed in the following sections.

\subsubsection{Querying the quadtree} \label{section:quadtree_querying}

\begin{figure}[h]
     \def\subfigwidth{.25\textwidth}
     \FIGURE
     {
          \subcaptionbox{\label{fig:qt_unresolved_edges_0}}
          {\includegraphics[width=\subfigwidth]{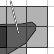}}
          \subcaptionbox{\label{fig:qt_unresolved_edges_1}}
          {\includegraphics[width=\subfigwidth]{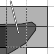}}
          \subcaptionbox{\label{fig:qt_unresolved_edges_2}}
          {\includegraphics[width=\subfigwidth]{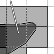}}
     }
     {Querying an edge through the quadtree: $\texttt{QT}(e)$. \label{fig:qt_unresolved_edges}}
     {(a) Edge $e$ is drawn in white. (b) The three relevant leaf nodes, highlighted with a dotted border. (c) $\texttt{QT}(e)$ returns indeterminable, the four unresolved edges ($E_{u}$), highlighted with a white border.}
\end{figure}

Figure \ref{fig:qt_unresolved_edges_0} shows an example of querying the quadtree with an edge: $\texttt{QT}(e)$.
The tree is traversed from the root node down in a depth-first manner.
Each node checks if the edge falls within its \textit{axis-aligned bounding box} (AABB).
If it does, the node is deemed \textit{relevant}.
Figure \ref{fig:qt_unresolved_edges_1} shows the relevant leaf nodes, two of which partially contain a hazard, while the topmost one (which is partially shown) is free of hazards.
The quadtree is, in this example, unable to definitively determine whether or not the edge collides with any of the hazards.

Despite this indeterminable outcome, substantial useful information is gained.
Since each node stores which edges from which hazard intersect, a set of unresolved edges $E_u$ can be derived from these relevant nodes.
Figure \ref{fig:qt_unresolved_edges_2} highlights the four unresolved edges, which will have to be resolved in the narrow phase (described in Section \ref{section:cde_edge_intersection}).

Whenever a single fully hazardous node is crossed, $\texttt{QT}$ returns \texttt{true} (collision).
If all crossed nodes are free of hazards, $\texttt{QT}$ returns \texttt{false} (no collision).
In all other cases, $\texttt{QT}$ returns \texttt{?} (indeterminable).

\subsubsection{Updating the quadtree} \label{section:quadtree_updating}
Any change with respect to the hazards (placing or removing an item) must also be reflected in the quadtree.
The quadtree must therefore be able to dynamically register and deregister hazards.
Updating a hazard entails traversing the tree and updating all affected nodes.
If a registration causes a leaf node ($<d$) to become partially hazardous, its four children will be generated.
Likewise, if a deregistration causes a node with children to become non- or fully hazardous, its children will be pruned.
In this way, the quadtree continuously adapts itself to its environment, increasing resolution where needed and reducing where it is not.

\begin{figure}[ht]
     \definecolor{white}{RGB}{255,255,255}
     \definecolor{gray}{RGB}{178,178,178}
     \definecolor{dark}{RGB}{54,54,54}

     \def\subfigwidth{.16\textwidth}
     \FIGURE
     {
          \subcaptionbox{\label{fig:qt_split_even}}
          {\includegraphics[width=\subfigwidth]{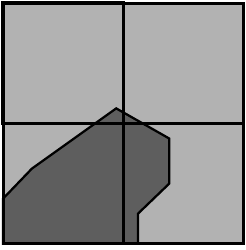}}
          \subcaptionbox{\label{fig:qt_split_free}}
          {\includegraphics[width=\subfigwidth]{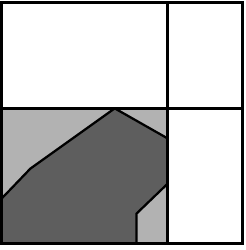}}
          \subcaptionbox{\label{fig:qt_split_full}}
          {\includegraphics[width=\subfigwidth]{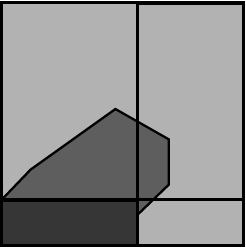}}
     }
     {Three ways to split a node in the quadtree. \label{fig:qt_split}}
     {Node contains $[$ \sqboxblack{white} no, \sqboxblack{gray} partial, \sqboxblack{dark} full $]$ hazard. 
     (a) Split in the center. 
     (b) Split to prioritize free nodes. 
     (c) Split to prioritize fully hazardous nodes.
     }
\end{figure}
When increased resolution is required in a particular node, it is split into four children.
The most obvious approach would be to split the node in the center (Figure \ref{fig:qt_split_even}).
The position of the split could also be determined dynamically, focussing on either generating non-hazardous (Figure \ref{fig:qt_split_free}) or fully hazardous nodes (Figure \ref{fig:qt_split_full}).
This would allow for more nodes to terminate at a shallower depth.
However, it also makes the tree structure heavily influenced by hazards that were (de)registered earlier on.
This bias could result in undesirable side effects when registering new hazards.
By contrast, splitting in the center allows the quadtree to remain entirely unbiased.
Since our CDE is operating in an optimization context where hazards are constantly being registered and deregistered, we opted for the center-split approach.

\subsection{Edge intersection} \label{section:cde_edge_intersection}

\begin{figure}[ht]
     \definecolor{darkgray}{RGB}{158,158,158}
     \definecolor{lightgray}{RGB}{235,235,235}
     \FIGURE
     {\includegraphics[width=0.7\linewidth]{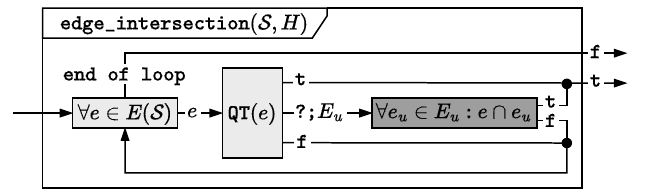}}
     {The \texttt{edge\_intersection} function. \label{fig:diagram_edge_intersection}}
     {[ \sqboxblack{lightgray} broad, \sqboxblack{darkgray} narrow ] phase, \texttt{t}: true, \texttt{f}: false, \texttt{?}: indeterminable.}
\end{figure}

The \texttt{edge\_intersection} function is the first of the two sub-procedures from the \texttt{collision} procedure illustrated in Figure \ref{fig:diagram_collision}.
This function will return \texttt{false} if shape $\mathcal{S}$ does not collide with any hazard by means of edge intersection.
Figure \ref{fig:diagram_edge_intersection} illustrates the procedural flow of the \texttt{edge\_intersection} function.

Every edge $e \in E(\mathcal{S})$ is tested by first querying the quadtree: $\texttt{QT}(e)$.
If the quadtree detects a collision (\texttt{t}), \texttt{edge\_intersection} immediately returns \texttt{true}.
If the quadtree concludes that no collisions occur (\texttt{f}), the function proceeds on to the next edge.
If the quadtree is undecided (\texttt{?}), the function returns a set of unresolved edges $E_u$ as described in Section \ref{section:quadtree_querying}.
A definitive answer is then produced during the narrow phase by performing line segment intersection tests for every edge $e_u \in E_u$.
Again, whenever a collision is detected, the procedure immediately returns \texttt{true}.
If all edges are cleared, the procedure returns \texttt{false}.

It is worth noting that the function does not strictly detect intersections between edges, but rather collisions between edges and the quadtree's nodes.
Once edge $e$ crosses a single fully hazardous node, the procedure will immediately detect the collision, regardless of whether $e$ is actually intersecting any edges.
If sufficient resolution (depth) is available in the quadtree, collisions caused by inclusion will, in many cases, already be detected here.
Therefore, the \texttt{edge\_intersection} function can detect a broader range of collisions than its name implies.

\subsection{Inclusion} \label{section:cde_inclusion}

Whenever \texttt{edge\_intersection} does not detect a collision, the \texttt{inclusion} procedure is called to eliminate any possibility of a collision by inclusion (Figure \ref{fig:diagram_collision}).
Given that all collisions with partial overlap are guaranteed to be detected by an edge intersection test (Section \ref{section:polygon_collision}), the only remaining undetected collisions are those where one shape is fully contained within another.
Since we could be dealing with hundreds of distinct hazards, the naive implementation (Section \ref{section:polygon_collision}) can become time-consuming. A more sophisticated method is required.

\begin{figure}[ht]
     \definecolor{darkgray}{RGB}{158,158,158}
     \definecolor{lightgray}{RGB}{235,235,235}
     \FIGURE
     {\includegraphics[width=0.85\linewidth]{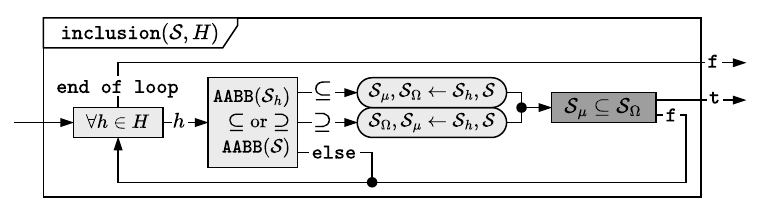}}
     {The \texttt{inclusion} function. \label{fig:diagram_inclusion}}
     {[ \sqboxblack{lightgray} broad, \sqboxblack{darkgray} narrow ] phase, \texttt{t}: true, \texttt{f}: false.}
\end{figure}

Figure \ref{fig:diagram_inclusion} illustrates the flow of the \texttt{inclusion} procedure.
Once again, performance is greatly improved by leveraging a two-phased approach.
The broad phase of the \texttt{inclusion} procedure is largely centered around the observation that, in order for one polygon to be fully contained in another, its respective axis-aligned bounding box (AABB) must also be fully contained in the AABB of the other:
\begin{equation}
     \mathcal{S}_\mu \subseteq \mathcal{S}_\Omega \implies \texttt{AABB}(\mathcal{S}_\mu) \subseteq \texttt{AABB}(\mathcal{S}_\Omega)
\end{equation}
The procedure loops over all hazards and checks if any of their AABBs is fully contained within the AABB of $\mathcal{S}$ or vice versa.
If one of the AABBs is fully contained in the other, an inclusion could occur.
In this case, it is also known which shape assumes the role of $\mathcal{S}_\mu$ and $\mathcal{S}_\Omega$.
This enables us to proceed to the narrow phase, where a single point-in-polygon test from $\mathcal{S}_\mu$ with respect to $\mathcal{S}_\Omega$ is performed.

In theory, any point $p \in \mathcal{S}_\mu$ could be used for the point-in-polygon test.
However, in order to avoid any potential numerical issues, we should also ensure that $p$ is not close to the boundary of $\mathcal{S}_\mu$ or $\mathcal{S}_\Omega$.
The pole of inaccessibility (PoI) of a polygon, explained in the following section, satisfies both of these requirements and is therefore an ideal candidate.

\section{Fail-Fast Surrogate} \label{section:fail_fast_surrogate}
Recall that the CDE is targeted towards irregular C\&P problems, where the goal is to pack the items in a compact fashion (as per Figure \ref{fig:example_irr_cp}).
It is therefore safe to assume that hazards will also be located in close proximity.
When validating item placements, the probability of colliding and non-colliding outcomes will not be balanced, with the former greatly outnumbering the latter.

The \textit{fail-fast} principle was first applied by \citet{haralick1980increasing} in tree search algorithms. They beautifully captured the essence of this concept when they stated: \textit{``to succeed, try first where you are most likely to fail''.}
This section will explore how we can leverage this principle to resolve \textit{obviously} colliding queries with as little effort as possible, thereby enhancing the overall efficiency of the CDE.

\begin{figure}[ht]
     \FIGURE
     {\includegraphics[width=0.45\textwidth]{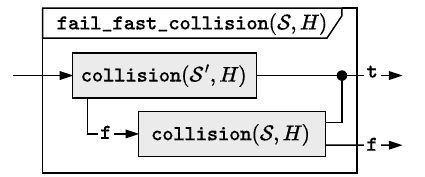}}
     {The \texttt{fail\_fast\_collision} procedure. \label{fig:diagram_fail_fast}}
     {A collision check on a surrogate shape $\mathcal{S'}$ precedes a check on the actual shape $\mathcal{S}$.}
\end{figure}

Figure \ref{fig:diagram_fail_fast} provides an abstract overview of the fail-fast approach applied to collision detection.
Prior to using the actual shape $\mathcal{S}$, collisions are first detected using a surrogate representation $\mathcal{S'}$.
When the surrogate is a subset of the original shape ($\mathcal{S'} \subseteq \mathcal{S}$), a colliding surrogate implies a collision of the original:
\begin{equation}
     \mathcal{S'} \cap \mathcal{S}_H \neq \emptyset \implies \mathcal{S} \cap \mathcal{S}_H \neq \emptyset
\end{equation}

A complete check using the actual shape is therefore only necessary when the algorithm detects no collision with the surrogate.
The surrogate model should be designed in a way to maximize the probability of detecting a collision with as little computational effort as possible.
For non-colliding queries, this additional check of the surrogate will slightly impair performance.
Nevertheless, given the aforementioned imbalance concerning the outcome of collision queries, this trade-off is usually worthwhile.

Converting polygons into a raster of pixels (\citet{segenreich1986optimal}), circle covering (\citet{rocha2013circle}) and semi-discrete representations (\citet{ma2007fast}) are three other examples of surrogate representations aimed to address the geometric challenge.
The difference between these approaches and our intended use is that they employ the surrogate as a total replacement for the original shape.
To ensure feasibility, these \textit{replacing} surrogates must cover all collision cases of the actual shape.
Consequently, the shape replacing a hazard must be a superset of the original hazard ($\mathcal{S'}_h \supseteq \mathcal{S}_h$).
This results in an inflation of the items and a deflation of the containers, which in turn causes a loss in precision (and maximum solution quality).
By instead using the surrogate merely as a preceding check (fail-fast), this significant drawback does not apply.

The following sections detail our implementation of a fail-fast surrogate, consisting of two distinct components.

\begin{figure}[ht]
     \def\subfigwidth{.25\textwidth}
     \FIGURE
     {
          \subcaptionbox{\label{fig:fail_fast_poles}}
          {\includegraphics[width=\subfigwidth]{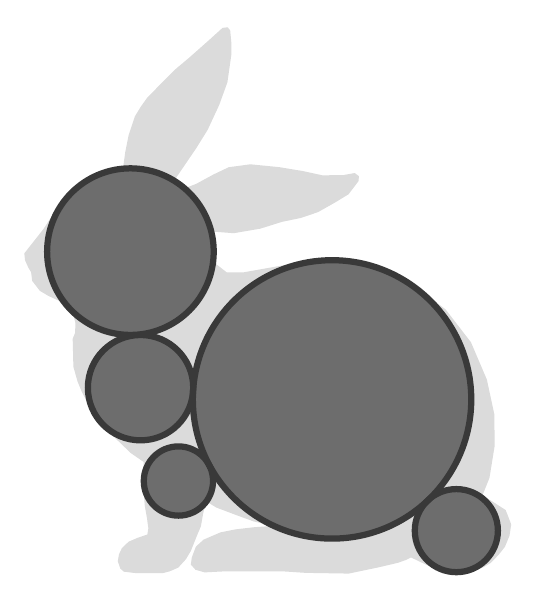}}
          \subcaptionbox{\label{fig:fail_fast_piers}}
          {\includegraphics[width=\subfigwidth]{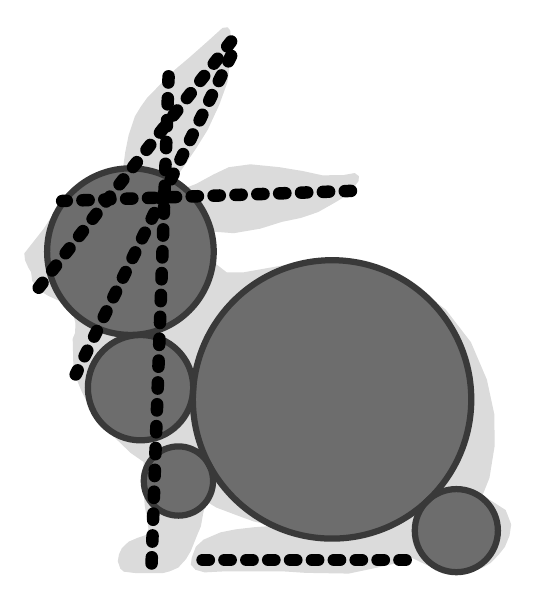}}
     }
     {Example of a fail-fast surrogate. \label{fig:fail_fast}}
     {(a) Poles. (b) Poles \& piers.}
\end{figure}

\subsection{Poles (of inaccessibility)} \label{section:poles}
The first component of the surrogate model involves a set of fully contained circles.
Figure \ref{fig:fail_fast_poles} shows an example where five such circles are inscribed within a bunny-shaped polygon $\mathcal{S}$.
A collision between any of these circles and a hazard also implies a collision with the original $\mathcal{S}$.
Detecting collisions using circles has three distinct advantages over line segments: they are (i) cheaper to transform, (ii) cheaper to query the quadtree with and (iii) more likely to collide.
Figure \ref{fig:fail_fast_poles} makes clear how just a few of these circles already cover the bunny relatively well.

We employ a concept that originates from geography to generate these circles. The Pole of Inaccessibility (PoI) defines the point in the interior of a shape that is the furthest away from its boundary.
It is important to note that the PoI is not synonymous with the centroid of a polygon.
A key distinction lies in the fact that, for non-convex polygons, the centroid is not guaranteed to be located in the interior of the shape.
From this point on, whenever we refer to the pole, we are referring to a circle with the PoI as its origin and the distance from the PoI to the boundary as its radius.

\citet{Agafonkin_Polylabel_a_fast_2016} proposed an efficient algorithm to calculate a polygon's PoI up to a certain (configurable) precision.
For use in the proposed surrogate, we extended their algorithm so that it can iteratively generate multiple non-overlapping poles.
Each iteration generates the next pole, with all poles from previous iterations being treated as holes in the shape. Figure \ref{fig:fail_fast_poles} shows the first five poles the algorithm generates for the bunny.

\subsection{Piers} \label{section:piers}
The main body of the bunny in Figure \ref{fig:fail_fast_poles} is captured quite well by the five poles, but its ears and legs are not represented at all.
Long and narrow sections are particularly challenging to capture with a limited set of circles.
This is not ideal since these `extremities' are quite likely to collide with hazards.
To combat this, we include a second component of surrogate representation to complement the poles.

Figure \ref{fig:fail_fast_piers} shows a set of (dashed) lines crossing the bunny, which we will refer to as \textit{piers}.
A pier is a line segment which is fully contained within the original shape, and selected in such a way to maximally complement the aforementioned limitations of the poles.
Figure \ref{fig:fail_fast_piers} shows that, with addition of a few piers, the shape of the bunny is now almost entirely captured.
The surrogate with both poles and piers enables us to detect virtually all `obvious' collisions, while being much cheaper to query than the actual shape.

\section{Polygon simplification} \label{section:polygon_simplification}

Despite the numerous efficiency enhancements introduced in the previous sections, 
both the query- and update-speed of the CDE remain affected by the number of edges in the polygons.
This sensitivity is not unique to 2D irregular C\&P problems and is often addressed by applying polygon simplification: the act of reducing the number of edges in a polygon while remaining as faithful to the original shape as possible.

For use in the CDE, simplified polygons are effectively a replacing surrogate (like those described in Section \ref{section:fail_fast_surrogate}) and must therefore cover all collision cases of the original shape.
In order to guarantee feasibility, original hazards have to be fully captured by their simplified versions:
\begin{equation}
     \mathcal{S}_h \subseteq \mathcal{S}'_h
\end{equation}
This means that entities are only allowed to be inflated or deflated depending on whether they induce an interior or exterior hazard, respectively:
\begin{equation}
     \begin{aligned}
          \mathcal{S}_e &
          \begin{array}{l}
               \left\{
               \begin{array}{lll}
                    \subseteq \mathcal{S'}_e & \text{if entity $e$ induces an interior hazard}
                    \\
                    \supseteq \mathcal{S'}_e & \text{if entity $e$ induces an exterior hazard}
               \end{array}
               \right.
          \end{array}
     \end{aligned}
\end{equation}
The procedure has to ensure that, when a simplification induces a change on the shape, it does so \textbf{strictly} outwards or inwards with respect to the original.

This requirement is generally not of concern for other applications of polygon simplification, which is why we cannot simply use any existing implementation such as \citet{douglas1973algorithms,fisher2005shape, visvalingam1993line} or \citet{wang1998line}.
Additionally, inflating hazards effectively restricts the search space and thus has a negative impact on the achievable solution quality.
Therefore, we do not wish to simplify the polygons in any way that would significantly alter them.
We are, in effect, seeking only to eliminate \textit{insignificant} features.

For purely academic optimization problems, such as the 2D nesting instances introduced by \citet{oliveira2000topos}, a simplification procedure may not be particularly effective because of the relatively simple shapes which have most likely been plotted manually.
However, real-world problems, such as the leather nesting dataset introduced by \citet{baldacci2014algorithms}, sometimes contain more detailed shapes generated by scans and/or arc approximations.
The simplification potential for these shapes is significant with almost no associated change in area.

\begin{figure}[ht]
     \def\subfigwidth{.32\textwidth}
     \def\subfigheight{.18\textheight}
     \FIGURE
     {
          \subcaptionbox{\label{fig:simpl_concave}}
          {\quad\quad\includegraphics[height=\subfigheight]{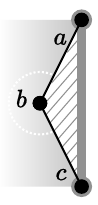}\quad\quad}
          \subcaptionbox{\label{fig:simpl_collinear}}
          {\quad\quad\includegraphics[height=\subfigheight]{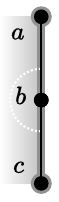}\quad\quad}
          \subcaptionbox{\label{fig:simpl_convexconvex}}
          {\quad\quad\includegraphics[height=\subfigheight]{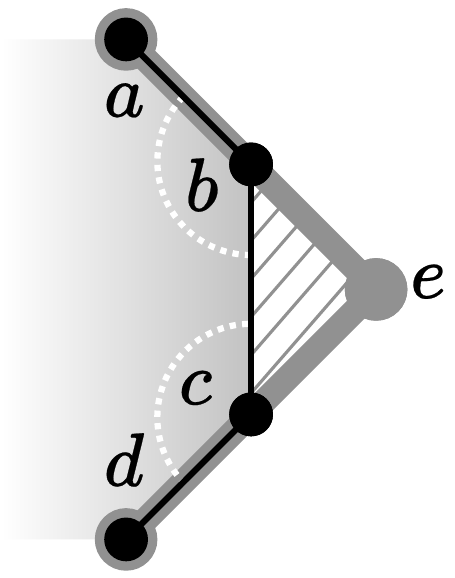}\quad\quad}
     }
     {Three options for simplification. \label{fig:polygon_simplification}}
     {Simplification by inflation illustrated. The angles are viewed from the interior of the polygon (shaded area). (a) Concave angle $\{a, b, c\} \rightarrow \{a, c\}$. 
     (b) Collinear angle $\{a, b, c\} \rightarrow \{a, c\}$. 
     (c) Consecutive convex angles $\{a,b,c,d\} \rightarrow \{a,e,d\}$.}
\end{figure}

Our approach consists of three elementary operations shown in the example of Figure \ref{fig:polygon_simplification}, in which a polygon is being simplified by inflation ($\mathcal{S'} \supseteq \mathcal{S}$).
The shaded area represents the interior of the polygon, while the hatched area denotes what is gained through the various simplification procedures.
The original vertices and edges are drawn in black, with the replacement ones in gray.
Each of these simplifications reduces the total number of edges by one.
The same holds for simplification by deflation if one simply reverses the interior and exterior of the polygon.

The simplification procedure is performed in an iterative and greedy way.
Each iteration generates all valid simplifications and calculates their associated differences in area.
A proposed simplification is rendered invalid if the resulting polygon would intersect itself.
The simplification with the smallest area change is selected and applied, each time reducing the number of edges by one.
Evidently, simplifications of a collinear angle are always eliminated first since they do not cause any inflation or deflation.
This process continues until no further simplification is possible without exceeding a configurable inflation/deflation threshold ($\alpha$).
This threshold is expressed as a percentage of the area of the original polygon, and is inversely correlated with the attainable compactness of the solution.

\begin{figure}[ht]
     \definecolor{darkgray}{RGB}{67,67,67}
     \definecolor{lightgray}{RGB}{202,202,202}
     \def\subfigwidth{.30\textwidth}
     \FIGURE
     {
          \subcaptionbox{Original, $|E| = 107$ \label{fig:poly_simpl_0}}
          {\includegraphics[width=\subfigwidth, angle=-90]{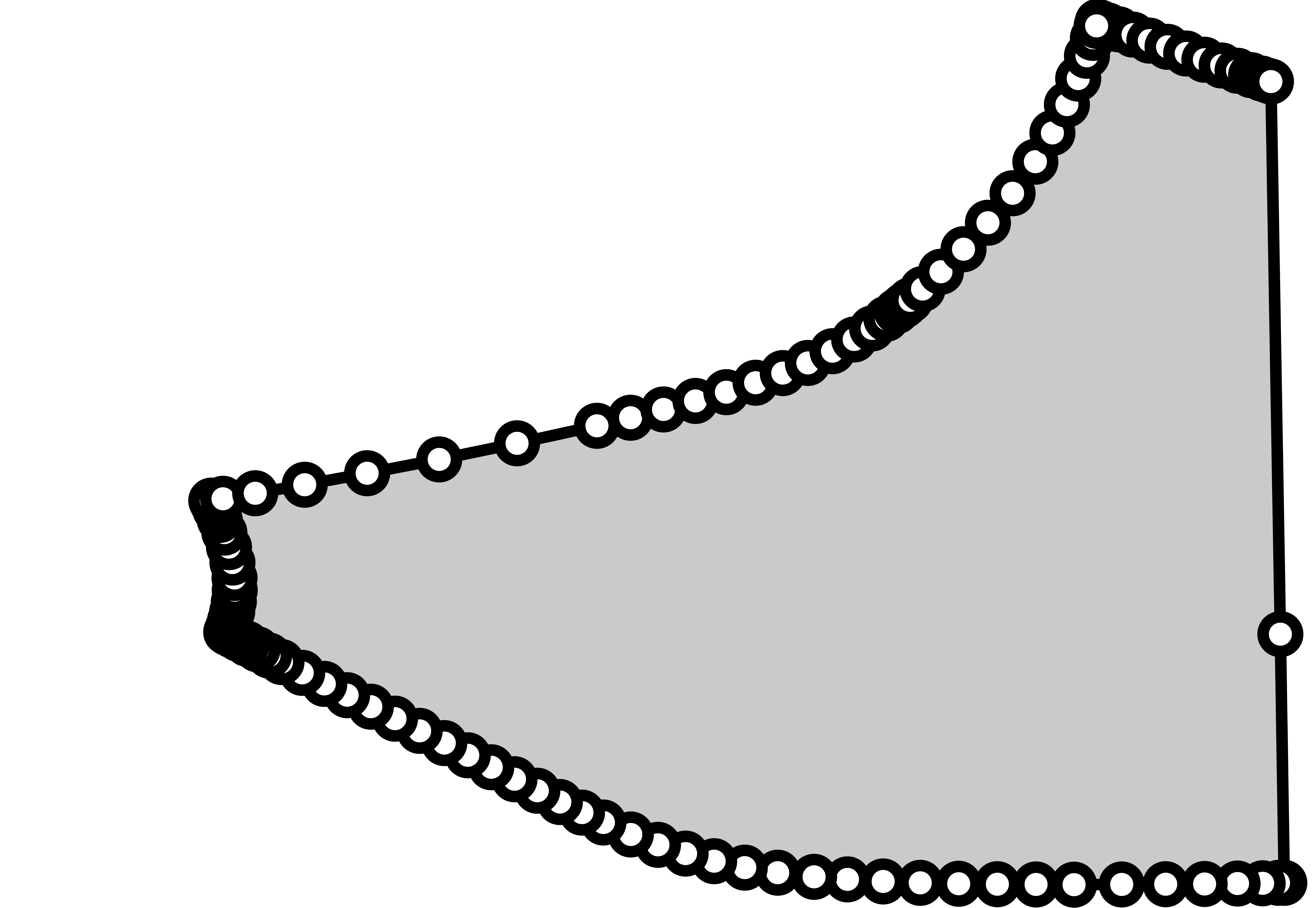}}
          \subcaptionbox{$\alpha = 0.1\%, |E'| = 36$ \label{fig:poly_simpl_1}}
          {\includegraphics[width=\subfigwidth, angle=-90]{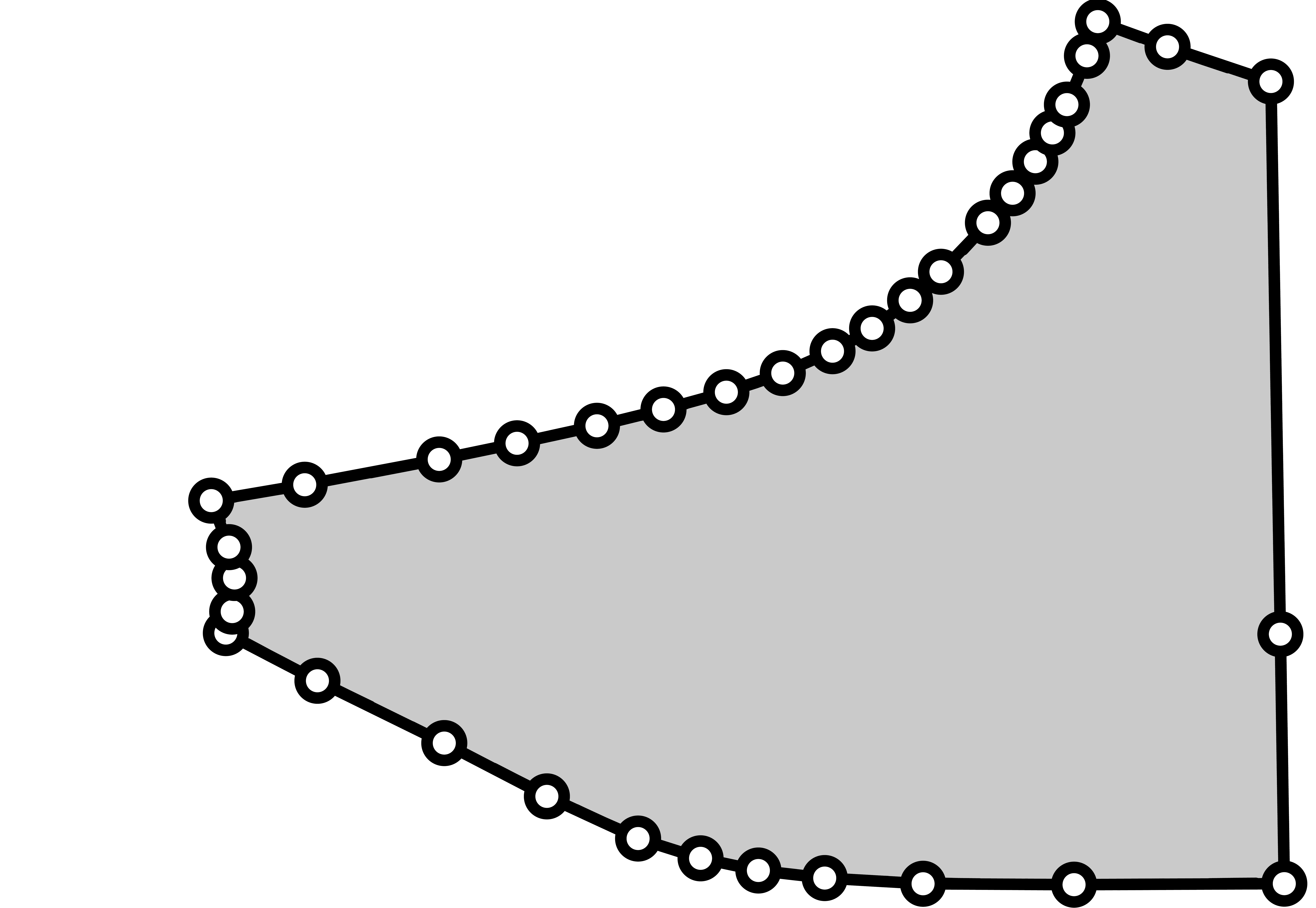}}
          \subcaptionbox{$\alpha = 1\%, |E'| = 13$ \label{fig:poly_simpl_2}}
          {\includegraphics[width=\subfigwidth, angle=-90]{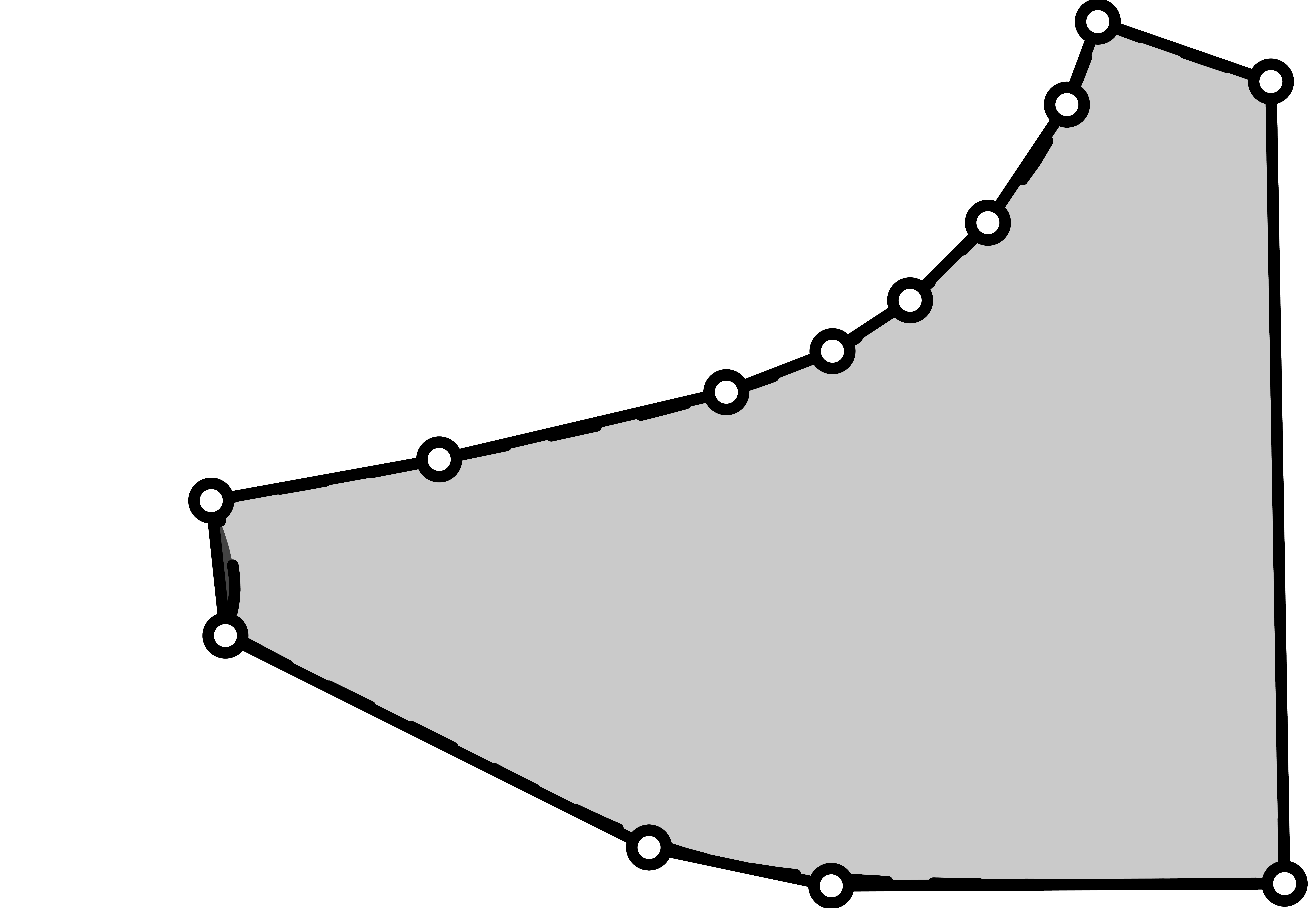}}
          \subcaptionbox{$\alpha = 10\%, |E'| = 6$ \label{fig:poly_simpl_3}}
          {\includegraphics[width=\subfigwidth, angle=-90]{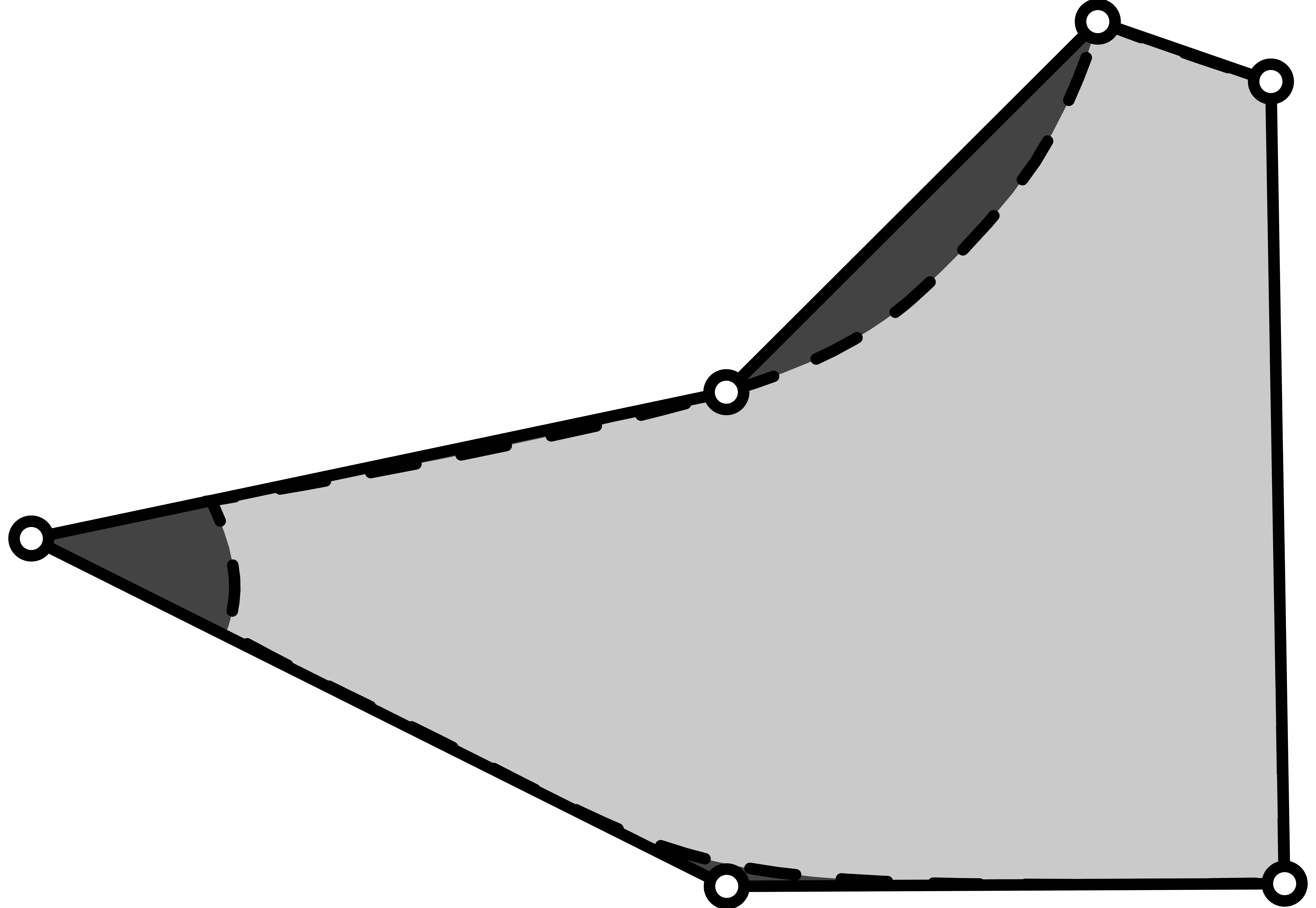}}
     }
     {Example of polygon simplification for different values of $\alpha$. \label{fig:poly_simpl_example}}
     {[\sqboxblack{lightgray} original, \sqboxblack{darkgray} simplified] shape, $ \mathcal{S}' \supseteq \mathcal{S}$, $|E'|$ denotes the remaining number of edges.}
\end{figure}

Figure \ref{fig:poly_simpl_example} illustrates our approach, simplifying a shape from the leather nesting dataset introduced by \citet{baldacci2014algorithms}.
In Figure \ref{fig:poly_simpl_1} and \ref{fig:poly_simpl_2}, the number of edges reduces by 66\% and 88\% respectively, with the simplified shape very closely resembling the original from Figure \ref{fig:poly_simpl_0}.
For more extreme simplifications, such as in Figure \ref{fig:poly_simpl_3} where 94\% of the edges are eliminated, the proposed approach is no longer very effective.
However, in an optimization setting, we are only interested in eliminating largely redundant features, and raising $\alpha$ to 1\% would already be too significant to be practical in many cases.

\section{Codebase}\label{section:codebase}

The CDE implementation is called \texttt{jagua-rs} and its source code is available at: \href{https://github.com/JeroenGar/jagua-rs}{\texttt{https://github.com/JeroenGar/jagua-rs}}, with a snapshot archived in the INFORMS Journal on Computing repository \citep{jaguars_repo}.
We wrote this project entirely in Rust: a statically-typed and compiled programming language with a strong emphasis on performance, safety and robustness.
We have currently implemented both strip- and bin-packing problems, however the project can be extended to support virtually any 2D irregular C\&P problem variant.

Given that the CDE operates in a continuous space, it is important to be aware of the inherent limitations of floating-point arithmetic.
Throughout our implementation, we have taken great care to ensure that false negatives --- undetected collisions --- do not occur.
For the polygon inclusion test from Section \ref{section:cde_inclusion} for example, we eliminate numerical issues that could arise when using a point that is close to the boundary of the shape by always opting to use the PoI.
When unable to eliminate, the design philosophy of \texttt{jagua-rs} is to always err on the side of caution.
Feasibility is guaranteed above all else, even if this means producing false positives in edge cases.
An important consequence of this philosophy is that for two shapes to be deemed non-colliding they must be strictly separated, even if only by an infinitesimal margin.
This limitation means that `exact fits' will always be deemed infeasible.
This is a deliberate choice and we consider it an acceptable trade-off to ensure the robustness and reliability of the CDE.

The repository also contains a basic optimization algorithm built on top of \texttt{jagua-rs}, namely a left-bottom-fill (LBF) heuristic.
This simple constructive heuristic selects the leftmost and bottommost position possible for each item.
It is far from the state of the art and simply serves as a reference implementation to demonstrate how researchers can couple their own optimization algorithms with \texttt{jagua-rs}.

\begin{figure}[ht]
     \FIGURE
     {\includegraphics[width=0.9\linewidth]{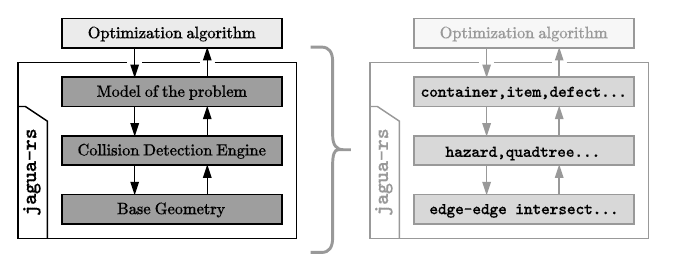}}
     {High-level overview of the different layers involved in solving a 2D irregular C\&P problem and the scope of \texttt{jagua-rs} \label{fig:diagram_jaguars_layers}}
     {The figure on the right lists some components discussed in previous sections, positioned at their corresponding layer.}
\end{figure}

Figure \ref{fig:diagram_jaguars_layers} provides a high-level visualization of the different layers involved in solving a 2D irregular C\&P problem and delimits the scope of \texttt{jagua-rs}.
The optimization algorithm --- which decides where to place items and which containers to use --- exists outside \texttt{jagua-rs}.
To give an impression of how someone might employ \texttt{jagua-rs} and where any modifications are required, we will briefly discuss three possible workflows.

First, imagine a researcher is developing a new optimization algorithm for the academic strip-packing problem.
This type of problem is already modeled in \texttt{jagua-rs} and therefore it can be used as a library (as-is), allowing the researcher to focus solely on their optimization algorithm without delving into any of the lower layers.

Second, consider a new real-world problem which features some additional constraints such as the quality zones explained at the end of Section \ref{section:hazards}.
In this case, not only does an optimization algorithm need to be developed, but \texttt{jagua-rs} also needs to be extended to model the additional spatial constraints (quality zones in this example).
Thanks to the fact that the CDE operates on hazards, which are general enough to express quality zones, these modifications are restricted to the top two layers and do not leak into the bottom two.

Third, there are those interested in improving the CDE itself and are not necessarily focused on an optimization challenge.
The decoupling illustrated in Figure \ref{fig:diagram_jaguars_layers} means their contributions to the bottom layers can be readily exploited by the previous two workflows without significant effort or interaction.

\section{Validation \& evaluation} \label{section:experiments}
While alternative approaches for collision detection in 2D irregular C\&P do exist (Section \ref{section:general_approaches}), no decoupled engine is currently available.
Even if one were, a direct and meaningful comparison would only be possible if the fundamental interactions (Section \ref{section:collision_detection_engine}) and limitations (Section \ref{section:general_approaches}) are similar.

In our opinion, the only irrefutable way to validate (i) the use of trigonometry, (ii) the general design of the CDE and (iii) its implementation, is to build a competitive optimization algorithm on top of \texttt{jagua-rs}.
Readers interested in seeing \texttt{jagua-rs} in action are referred to the \texttt{sparrow} project: an open-source heuristic for the 2D strip packing problem (\citet{gardeyn2025ispp}).
Built on top of \texttt{jagua-rs}, \texttt{sparrow} demonstrates the CDE's capabilities in practice.

For readers interested in a relative performance evaluation, we use the \texttt{criterion.rs} framework by \citet{criterion} to perform “statistics-driven microbenchmarking” of the most critical operations within the CDE.
We encourage readers interested in using the CDE to re-run the tests on their own hardware and problem instances using the latest version of \texttt{jagua-rs}.
These benchmarks should provide valuable insights into the factors at play and the trade-offs involved when tuning the CDE for maximum performance.

\section{Conclusion}
There are two challenges involved in any irregular cutting and packing problem.
First, there is the geometric difficulty of ensuring items are placed in such a way that they do not collide with each other, with the boundary of the container or certain restricted regions within.
Second, there is the optimization challenge of finding a good solution in accordance with some objective.
Regardless of the method used to reach a high-quality solution, being able to quickly and accurately determine its geometric feasibility is of universal importance.
Until now, there has not been a general-purpose and reusable way of handling this shared geometric challenge.

The collision detection engine (CDE) we have introduced in this paper addresses this gap by decoupling geometry from optimization.
The CDE provides researchers and developers with a powerful and adaptable tool that can quickly determine whether or not a certain placement of items is feasible. This enables them to confidently focus their efforts on the optimization challenge at hand, without worrying about the underlying geometry.

Our paper introduced a suite of core principles and design philosophies which we believe are essential when modeling a CDE for use in irregular C\&P problems: \textit{hazards, two-phased approach, fail-fast surrogate} and \textit{polygon simplification}.
However, such concepts are only useful insofar as they can actually be leveraged to achieve an effective implementation.

Therefore, we also shared one possible interpretation of these concepts which, when combined, resulted in a concrete implementation. 
\texttt{jagua-rs} targets maximum performance and robustness, being able to resolve millions of queries per second.
The source code for this project is available at: \href{https://github.com/JeroenGar/jagua-rs}{\texttt{https://github.com/JeroenGar/jagua-rs}}

We intend for this paper and \texttt{jagua-rs} to serve as catalysts for further research and development in the field of irregular C\&P problems, providing solid foundations on which to build.
While readers could certainly directly reuse or adapt \texttt{jagua-rs}, we would also encourage further experimentation and interpretation of these building blocks to arrive at something that works best for their particular context.

\ACKNOWLEDGMENT{
The authors wish to explicitly thank Luke Connolly (Connolly Editorial) for his invaluable editorial consultation, Patrick De Causmaecker (KU Leuven) for help with the mathematical notations and Alessio La Greca (KU Leuven) for meticulously proofreading the final version of the paper.
The authors also thank the editors and reviewers, whose efforts have undoubtedly improved the clarity of the paper.
}

\bibliographystyle{informs2014}
\bibliography{bibliography}

\end{document}